\documentclass[prl,a4paper,aps,twocolumn,showpacs,superscriptaddress,groupedaddress]{revtex4}
\usepackage[dvips]{graphicx}
\usepackage{ae}
\usepackage[T1]{fontenc}
\usepackage[ansinew]{inputenc}
\usepackage{amsmath}
\usepackage{amssymb}
\usepackage{graphicx}
\usepackage{color}
\usepackage[colorlinks]{hyperref}
\usepackage{lscape}
\begin{document}
\title{Simulating GHZ Correlations Relaxing Physical Constraints}
\author{Biswajit Paul}
\email{biswajitpaul4@gmail.com}
\affiliation{Department of Mathematics, St.Thomas' College of Engineering and Technology, 4,Diamond Harbour Road, Alipore, Kolkata-700023, India.}
\author{Kaushiki Mukherjee}
\email{kaushiki_mukherjee@rediffmail.com}
\affiliation{Department of Applied Mathematics, University of Calcutta, 92, A.P.C. Road, Kolkata-700009, India.}
\author{Debasis Sarkar}
\email{dsappmath@caluniv.ac.in}
\affiliation{Department of Applied Mathematics, University of Calcutta, 92, A.P.C. Road, Kolkata-700009, India.}

\begin{abstract}
Violation of Bell inequality (or, Bell-type inequalities) by nonlocal correlations is justified by relaxation of at least one of the plausible physical constraints used to model such inequality. Based on this fact, in this letter we present a procedure to simulate three-qubit GHZ correlation relaxing two constraints, determinism and no signaling simultaneously. We have also derived the minimum amount of indeterminism and signaling to be introduced in a system. The corresponding number of signaling and local bits of mutual information needed to communicate are also provided and thus we are able to focus on utility of relaxation of these two constraints as useful resources.
\end{abstract}
\date{\today}
\maketitle

In any experiment on a model consisting of at least two subsystems, the statistics of the experimental results develop ceratin correlations. When the model is based on quantum theory then in some cases the correlations simulated cannot be obtained by any local theory. This type of correlations are referred as \textit{non-local correlations}. Non-local correlations show deviations from correlations that are developed in a local model. One can manifest such correlations  by the violation of Bell inequality \cite{Bell}. This inequality with the modifications known as Bell-type inequalities, are capable of distinguishing local and non-local correlations. Violation of Bell-type inequalities by predictions made in quantum theory implies that this theory is capable of simulating non-local correlations.

Several attempts made recently to simulate non-local correlations. The study of non-local correlations can be broadly classified into four distinct categories. One category frames non-local correlations as a resource in information processing; e.g., random number generators \cite{Pir}, device independent quantum key distribution \cite{Bar, Acin, Pir1}, etc. Second category of study focusses on the question: "why quantum theory is not more non-local" \cite{Pop}. Third group of researchers are interested to measure non-locality. They have quantified non-locality as the minimal number of classical bits that are needed to communicate from one party to another in order to simulate the required correlation \cite{Mau, Bra, Ste, Gis}. For example, Toner and Bacon \cite{Bac} gave a protocol where one bit of communication was sufficient to generate singlet correlations(considering Von-Neumann measurements). In \cite{Bbent}, it was shown that for classical simulation of an n-party GHZ state, at least $n\log_{2}n\,-2n$ bits of communication is necessary. Using single N-qubit entangling pulse in a network (fully connected) of qubits interacting by anisotropic Heisenberg exchange, one can also simulate GHZ correlations \cite{Ab}. In \cite{Nai}, quantum mechanical predictions for measurements of arbitrary products of Pauli operators on GHZ state were generated using a local hidden variable model in which number of classical bits that was required to be communicated, varied linearly with the number of qubits used. N. Gisin \cite{Bran} introduces a protocol that reproduces three-partite GHZ correlations with bounded communications. To be more specific, it was shown that total three bits of classical communications were sufficient for generation of all equatorial Von-Neumann measurements on the three-partite GHZ state.

The Bell-type inequalities together with the original Bell inequality, were derived under the basic assumption that these inequalities abide by some physical constraints namely no signaling conditions, free will, determinism, etc. M. J. W. Hall argued that the violation of any of these inequalities by any consistent theory such as quantum mechanics can be justified by relaxing one or more than one of such physical constraints at a time\cite{Hall1, Hall2, Hall3}. The fourth group of study deals with the simulation of singlet correlations with the aid of \textit{relaxed Bell inequalities}\cite{Hall1, Hall2, Hall3}. In \cite{Hall1}, minimal degree of relaxation of measurement independence for generating singlet correlations is derived and in \cite{Hall2}, the minimal degree of joint relaxation of no signaling and determinism for the same is considered. Lastly, in \cite{Hall3}, we find simultaneous relaxation of measurement independence, no signaling and determinism. Till date relaxation of physical constraints has been used to simulate two party non local correlations only. Therefore, one can naturally ask the question of simulating the same for more than two parties. In this letter, we have tried to find a solution to this problem. Our work basically belongs to the last two groups of study. Here we have simulated three-qubit GHZ \cite{GHZ} correlations using joint relaxation of no signaling and determinism as resources and we obtain the minimal number of signaling bits of mutual information that is needed to be communicated for this purpose. We have considered two scenarios of relaxations in this context: \textit{Simultaneous Relaxation Scenario} and \textit{Restricted Relaxation Scenario}.

Consider three parties Alice, Bob and Charlie. Any two of them are together and form a group (say, $G_1$) while the other one remains separated and forms a second group (say, $G_2$). In the first scenario, signaling can take place in all possible directions between $G_1$ and $G_2$. But in the second scenario, signaling is restricted to communications in some possible directions between the two groups only. Without loss of generality, we consider two subcases in this scenario : (i) Charlie sends a signal either one /or to both of Alice and Bob; (ii) Charlie receives a signal to either one /or from both of Alice and Bob.  This scenario is thus a special case of the previous scenario as any one of the three parties either sends a signal to (first subcase) or receives a signal from (second subcase) either one/or both of the remaining parties at a time.

Now, the degree of signaling is defined as the maximum shift possible in an underlying marginal probability distribution for one group, due to the alteration of measurement settings by the other. It may be formulated as \cite{Hall2}:
\begin{widetext}
\begin{equation}\label{s1}
    S_{i\rightarrow jk}\,:=\,\sup_{x_i,x_i',x_j,x_k,a_j,a_k,\lambda}|\,p^{(jk)}(a_j,\,a_k|\,x_i,\,x_j,\,x_k,\,\lambda)\,-\,p^{(jk)}(a_j,\,a_k|x_i',\,x_j,\,x_k,\,\lambda)\,|,
\end{equation}
\begin{equation}\label{s2}
    S_{ij\rightarrow k}\,:=\,\sup_{x_i,x_i',x_j,x_j',x_k,a_k,\lambda}|\,p^{(k)}(a_k|\,x_i,\,x_j,\,x_k,\,\lambda)\,-\,p^{(k)}(a_k|x_i',\,x_j',\,x_k,\,\lambda)\,|,
\end{equation}
\qquad\qquad\qquad\qquad\qquad\qquad\qquad\qquad\qquad\qquad\qquad\qquad\qquad$\forall\, i,j,k\,\in\{1,2,3\}$ and $i\neq j\neq k$.
\end{widetext}
where $x_i\in \{0,\,1\}, \,\forall i$ and $a_i\in \{-1,\,1\}, \,\forall i$ denotes the measurement settings  and outcomes of Alice (for, $i$=1), Bob (for, $i$=2) and Charlie (for, $i$=3), respectively. Clearly, $ S_{ij\rightarrow k},\,S_{i\rightarrow jk}\geq \, 0$, $\forall i,j,k\,\in\{1,2,3\}\,\text{and}\, \: i\neq j\neq k$. For $S_{ij\rightarrow k}> \,0$, group $G_1$ sends a signal to $G_2$ and $G_2$ sends signal to $G_1$ if $S_{i\rightarrow jk}> \,0$. The overall degree of signaling, for the whole system is defined as,
\begin{equation}\label{s3}
    S:=\max_{i,j,k}\,\{S_{i\rightarrow jk},\,S_{ij\rightarrow k}\},~\forall i,j,k \in\{1,2,3\}\, \text{and} \,\: i\neq j\neq k.
\end{equation}

The degree of indeterminism of an underlying model may be defined as the measure of deviation of the marginal probabilities from the deterministic values 0 and 1. The local degrees of indeterminism may be defined as the smallest positive numbers $I_{ij}$ and $I_{i}$ such that the corresponding marginal probabilities lie in $[0,\,I_{ij}]\,\bigcup\,[1-I_{ij},1]$ and $[0,\,I_{i}]\,\bigcup\,[1-I_{i},1]$ respectively \cite{Hall2}, i.e.,

\vspace{.5cm}

 $I_{ij}\,:=$
 \begin{equation}\label{s4}
     \sup_{\{x_i,x_j,\lambda\}}\min_{\{a_i,a_j\}}\{p^{(ij)}(a_i,a_j|x_i,x_j,\lambda), 1-p^{(ij)}(a_i,a_j|x_i,x_j,\lambda)\},
\end{equation}
$I_{i}\,:=$
\begin{equation}\label{s5}
    \sup_{\{x_i,\lambda\}}\,\min_{\{a_i\}}\{p^{(i)}(a_i|\,x_i,\,\lambda),\,1-\,p^{(i)}(a_i|\,x_i,\,\lambda)\},
\end{equation}

\hspace{3cm}$\forall\, i,j\,\in\{1,2,3\}$ and $i\neq j$.

\vspace{.5cm}

Hence $I_{ij}$ (or, $I_i$) = 0 if and only if the corresponding marginal is deterministic. The overall degrees of indeterminism for the system may be defined as,
\begin{equation}\label{s6}
     I\,:=\,\max_{{i,j}}\{I_{ij},\,I_i\}, \, \forall\, i,j\,\in\{1,2,3\} \,\textmd{and}\: i\neq j.
 \end{equation}

\textit{Complementary Relation between Signaling and Indeterminism}: Due to signaling, any deviation in a marginal probability value $p^{(ij)}$ or, $p^{(i)}$($\forall i$) must either retain the value in the same subinterval $[0,\,I]$ (or $[1-I,1]$) for $S<1-2I$ or, shift it across the gap between the subintervals $(S\geq 1-2I)$. This provide us the relation: $I\geq \textmd{min}\{S,\,(1-S)/2\}$.\\

\textit{Svetlichny Inequality \cite{Svet}:} This Bell-type inequality (for three party correlations) can distinguish 2/1 party non-locality from three party non-locality.
 \begin{equation}\label{s7}
 \begin{split}
 &\langle X_1X_2X'_3\rangle\,+\,\langle X_1X'_2X_3\rangle\,+\,\langle X'_1X_2X_3\rangle\,+\,\langle X'_1X'_2X_3\rangle\\
 & \,+\,\langle X'_1X_2X'_3\rangle\,+\,\langle X_1X'_2X'_3\rangle\,-\,\langle X_1X_2X_3\rangle\,-\,\langle X'_1X'_2X'_3\rangle\\
 &\,\leq\,\mathcal{S}.
 \end{split}
 \end{equation}
where $\langle X_1X_2X_3\rangle$ has its usual meaning; $\mathcal{S}\leq4$ for 2/1 party non-local correlations; $8 \geq \mathcal{S}>4$ for three party non-locality. For GHZ correlations $\mathcal{S}=4\sqrt{2}$ \cite{Rob}. This violation of Svetlichny inequality is used to simulate GHZ correlations. For that we first sketch Svetlichny inequality in \textit{simultaneous relaxation scenario}.
\\\textit{Theorem 1}: For any underlying model having values of indeterminism and signaling of at most $I$ and $S$, $\mathcal{S}=\mathcal{S}(I,S)$, where, \textmd{for}\: $S<1-2I,$
\begin{equation}\label{s8}
\begin{split}
\quad\mathcal{S}(I,S)\,&=\,4+12I \,\quad\textmd{ for} \:0\,\leq \,I\,\leq\,2/9,\\
                       &=\,48I-4 \quad \,\textmd{for}\: 2/9\,\leq \,I\,<\,1/4,\\
                       &=\,8      \qquad\quad\quad\textmd{for}\:1/4\leq I\,\leq\,1/2,\\
\end{split}
\end{equation}
and \textmd{for}\: $S\geq1-2I$,
\begin{equation}\label{B2}
\quad\mathcal{S}(I,S)\,=\,8.
\end{equation}
Proof. See Appendix.

For deterministic no signaling model Eq.(\ref{s8}) and Eq.(\ref{B2}) reduce to the original Svetlichny inequality(\ref{s7}).\\
\textit{Simulation of GHZ correlations}: If $V$ denotes the amount of violation of Eq.(\ref{s7}), $\mathcal{S}(I,S)\geq4+V$. By Eq.(\ref{s8}), this results to
$I\,\geq\,I_V$ \textmd{and/or} $S\,\geq \,S_V$ \textmd{where}
\begin{equation}\label{s9}
\begin{split}
&I_V:= V/12;\,\, S_V:=1-V/6 \: \,\, \textmd{for} \: 0\,\leq \,I\,\leq\,2/9, \\
&I_V:=1/6+V/48;\,\, S_V:= 2/3-V/24 \:\,\,\textmd{for}\: 2/9\,\leq \,I<1/4.
\end{split}
\end{equation}
For GHZ correlations, $V=4\sqrt{2}- 4$ \cite{Rob}. Hence if $0\,\leq \,I\,\leq\,2/9$, at least 0.13 amount of indeterminism must be induced in the underlying marginal probability distribution of the system containing $G_1$ and $G_2$ and/or at least 0.72 amount of signaling must be communicated between $G_1$ and $G_2$; if $2/9\,\leq \,I\,<\,1/4$, at least 0.20 amount of indeterminism and/or at least 0.59 amount of signaling is required.

\textit{Quantifying Non-locality}: There is an information theoretic interpretation of the above equation (\ref{s9}). The number of non-local signaling bits of mutual information required to communicate between two groups to simulate GHZ correlations quantifies non-locality \cite{fuch}. From equations (4), (5), (6), we find there is an underlying probability distribution of marginals which is closed to the distribution $(I,1-I)$ and correspondingly the Shannon entropy which measures the content of information is close to $H(I)= -I\log_2 I -(1-I)\log_2 (1-I)$\cite{Hall2}. Again considering the equations (1), (2), (3), with the choice of one of the groups measurement procedures, the other one will obtain a marginal probability distribution $(p,1-p)$ or $(p+S, 1-p-S)$. If we consider both are equiprobable, then the mutual information that can be communicated between two groups is given by \cite{Hall2},
$$ J(p,p+S; 1/2, 1/2)= H(\frac{p}{2}+\frac{p+S}{2})- \frac{1}{2}(H(p)+H(p+S)),$$ where $J$ attains minimum for $p=\frac{1-S}{2}.$ Thus the capacity of a binary symmetric channel based on the maximum probability shift $S$, due to signaling is $C(S)$, given by $C(S) = 1- H(\frac{1-S}{2})$ \cite{Hall2} bits per joint measurement. So, by Eq.(\ref{s9}),
\begin{equation}\label{s10}
    \begin{split}
      C(S_V)\,=0.43 & \quad \textmd{for}\:0\,\leq \,I\,\leq\,2/9, \\
       =0.71 & \quad  \textmd{for}\: 2/9\,\leq \,I\,<\,1/4
    \end{split}
\end{equation}
where $C(S_V)$ is the channel capacity. This gives the minimal number of non-local signaling bits to be sent from one group to the other.
\\ Next we try to  generate GHZ correlations under \textit{restricted relaxation scenario}.\\
\emph{\textbf{1st sub case}}: Here Svetlichny inequality takes the form:
\\\textit{Theorem 2}: For $S_{3\rightarrow 12}\,>\,0;\:S_{i3\rightarrow k}>0\,(i,k\in \{1,2\};\,i\neq k)$, $\mathcal{S}=\mathcal{S}(I,S)$ gets modified as:
\begin{equation}\label{s12}
\begin{split}
\mathcal{S}(I,S)\,=\,4+8I &\quad  \textmd{for}\: S\,<\,1-2I\\
 =\,8 &\quad\textmd{for}\:S\,\geq\,1-2I\\
\end{split}
\end{equation}
Proof: From the restrictions imposed by the statement of the theorem, one obtain the tight upper bound,
\begin{equation}\label{A1}
\begin{split}
   J &\geq 2(1 - 2I) \qquad\qquad \text{for}\quad S < 1 - 2I\\
     &\geq 0  \qquad\qquad \qquad \quad \text{for} \quad S \geq 1 - 2I
\end{split}
\end{equation}
 Equality is obtained for $S < 1 - 2I$, for example, when $m_{i1} = m_{i3} = m_{i6} = 0$(for $i = 1, \ldots, 8$ ), $m_{i2} =0$ (for $i = 1, \ldots, 6$), $m_{i4} = 1 - m_{i5} = I$ (for $i = 1, \ldots, 6$), $m_{i2} = m_{i4} = 1 - m_{i5} = I$ (for $i = 7, 8$) and for $S\geq 1 - 2I$, when $m_{i1} = m_{i3} = m_{i6} = 0$(for $i = 1, \ldots, 8$ ), $m_{i2} =0$ (for $i = 1, \ldots, 6$), $m_{i4} = 1 - m_{i5} = I$ (for $i=1, \ldots, 6$), $m_{i2} = m_{i4} = m_{i5} = 0$ (for $i = 7, 8$).
Hence Eq.(\ref{s12}) follows by using (Eq.(\ref{A1}) and Eq.(\ref{Z3})). $\blacksquare$\\
\emph{\textbf{2nd sub case}}: We now frame Svetlichny inequality in this scenario as follows:
\\\textit{Theorem 3}: For $S_{12\rightarrow 3}\,>\,0;\:S_{i\rightarrow j3}>0\,(i,j\in\{1,2\};\,i\neq j)$, $\mathcal{S}=\mathcal{S}(I,S)$ gets modified as:
\begin{equation}\label{s11}
\begin{split}
\mathcal{S}(I,S)\,=\,4+4I &\quad  \textmd{for}\: S\,<\,1-2I\\
 =\,8 &\quad\textmd{for}\:S\,\geq\,1-2I\\
\end{split}
\end{equation}
Proof: This is analogous to the proof of \textit{Theorem 2}. $\blacksquare$

\textit{Comparison between different scenarios for $S\,<\,1-2I$ case}: $\mathcal{S}(I,S)$ for \textit{Simultaneous Relaxation Scenario} lies in $[4,\,6.66]$ for $0\leq I\leq\frac{2}{9}$; [6.66,\,8) for $ \frac{2}{9}\leq I < \frac{1}{4}$; $\{8\}$ for $\frac{1}{4}\leq I<\frac{1}{2}$ (see FIG. 1, where X represents $\mathcal{S}$ for simultaneous relaxation scenario). So maximum violation is obtained in this scenario for $S < 1 - 2I $.  But maximum violation is not obtained for \textit{Restricted Relaxation Scenario} where $\mathcal{S}$ lies in $[4, 8)$ and $[4,6)$ for \textbf{1st sub case} and \textbf{2nd sub case} respectively (see FIG. 1, where Y and Z represent $\mathcal{S}$ for these two subcases of Restricted relaxation scenario respectively).
%\begin{widetext}
\begin{figure}[htb]
\centering
\includegraphics[width=3in]{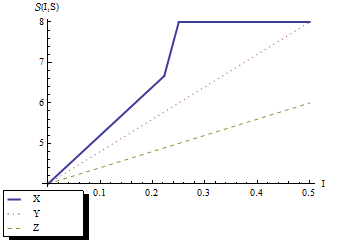}
\caption{\emph{The upper bounds $\mathcal{S}(I,S)$ for \textit{Simultaneous Relaxation Scenario} and \textit{Restricted Relaxation Scenario} (\textbf{1st sub case},\,\textbf{2nd sub case}) are plotted for $S\,<\,1-2I$ case. } }
\end{figure}
%\end{widetext}
\begin{widetext}
\begin{center}
\textit{Minimal Requirements for Simulation of GHZ Correlations}
\end{center}
\begin{tabular}{|c|c|c|c|c|}
  \hline
  Subcases & Minimal Requirement Of Relaxation ($I_V,S_V$) &Minimal Amount Of Relaxation & Signaling Bits & Local Bits  \\
  \hline
1& $I\geq I_V:= V/8$ and/or $S\geq S_V:=1-V/4$&$I_V=0.20\quad \textmd{and/or}\quad S_V=0.58$&0.27&0.73 \\

 \hline
2&$ I\geq I_V:= V/4\quad \textmd{and/or} \quad S_V:=1-V/2$&$I_V=0.41\quad \textmd{and/or}\quad S_V=0.17$&0.03&0.97\\
 \hline
 \end{tabular}
\end{widetext}

\textit{Conclusion}: From all those we have been discussed above, it can now safely be concluded that this work introduces a generalized procedure to simulate not only GHZ but any tripartite non-local  correlations using relaxation of two physical constraints of determinism and no signaling in a system. We have given the minimal degrees of joint relaxations of no signaling and determinism required for simulation of tripartite GHZ correlations. For this, we have considered two scenarios: one in which physical constraints are relaxed for all the three parties at a time and another scenario where partial relaxation takes place, i.e., constraints are not relaxed at least for one party at a time. We have also given a measure of non local bits (in terms of signaling bits of mutual information and also the corresponding number of local bits hence focusing on the trade-off between the two) required to be communicated  for this purpose. Apart from its importance as a possible way to generate GHZ correlations, this work may be considered as a step towards introducing procedures to simulate multi-party correlations via relaxation of physical constraints. Besides, in \textit{simultaneous relaxation scenario}, the maximal violation of any tripartite bell-type inequality is obtained in the region $S\,<\,1-2I$ (for $1/4\leq I<1/2$) unlike that in the bipartite case where maximal violation of CHSH inequality is obtained only when the marginals ($p$) lie across the gap ($S\,\leq\,1-2I$) \cite{Hall2}. But in \textit{restricted relaxation scenario}, maximal violation is obtained only across the gap. This is indeed a striking feature, the reason for which is not apparent and can be explored further.\\
{\bf Acknowledgement:} The authors thank Ajoy Sen for interesting and helpful discussions relating to the topic of this work. The author KM also acknowledges the financial support by University Grants Commission(UGC), New Delhi.

\appendix

\begin{widetext}

\begin{center}
{\bf Appendix: Proof of Theorem 1.}
\end{center}
The technique of proof is like in \cite{Hall2}.
Proof: To obtain the relations in Eqs. (\ref{s8}) and (\ref{B2}), in case of two valued measurement settings, denote the possible outcomes by $\pm 1,$ and the joint measurement outcomes are ordered as $(+,+,+)$, $(+,+,-)$, $(+,-,+)$, $(-,+,+)$, $(+,-,-)$, $(-,+,-)$, $(-,-,+)$ and $(-,-,-)$. Suppose, $P(+,+,+|x_1,x_2,x_3,\lambda) = c$, $P(+,+|x_2,x_3,\lambda) = m_1$, $P(+,+|x_1,x_2,\lambda) = m_2$, $P(+,+|x_1,x_3,\lambda) = m_3$, $P(+|x_1,\lambda) = m_4$, $P(+|x_2,\lambda) = m_5$ and $P(+|x_3,\lambda) = m_6$.
For a fixed value of $\lambda$, $\langle X_1X_2X_3 \rangle_{\lambda} = 8c -1 + 2(m_4 + m_5 + m_6) - 4(m_1 + m_2 + m_3)$. Restricting the range of $c$ by the positivity condition of probability, we get
%\begin{widetext}
\begin{equation}\label{Z1}
\begin{split}
    &-1 + 2\{|m_2 + m_3 - m_4| + |m_2 + m_6 - m_3 - m_5| + |m_5 + m_6 - m_4 -2m_1 + |m_2 + m_3 - m_4| - |m_2 + m_6 - m_3 - m_5||\}\\
    &\leq \langle X_1X_2X_3 \rangle_{\lambda} \leq  1 - 2 \{|m_1 - m_2| + |1 + m_1 + m_2 - m_4 - m_5 - m_6| + |m_4 + m_5 + m_6 - 2m_3 - 1 \\
    & -|m_1 - m_2| + |1 + m_1 + m_2 - m_4 - m_5 - m_6||\}
\end{split}
\end{equation}
where upper and lower bounds are attainable for suitable choices of c.
%\end{widetext}
Label the eight measurement settings $(x_1,x_2,x'_3,)$, $(x_1,x'_2,x_3,)$, $(x'_1,x_2,x_3,)$, $(x'_1,x'_2,x_3,)$, $(x'_1,x_2,x'_3,)$, $(x_1,x'_2,x'_3,)$, $(x_1,x_2,x_3,)$ and $(x'_1,x'_2,x'_3,)$, by $1$, $2$, $3$, $4$, $5$, $6$, $7$ and $8$ respectively and define,
\begin{equation}\label{Z2}
\begin{split}
 & E_{\lambda} =  \langle X_1X_2X'_3\rangle\,+\,\langle X_1X'_2X_3\rangle\,+\,\langle X'_1X_2X_3\rangle\,+\,\langle X'_1X'_2X_3\rangle\\
 & \,+\,\langle X'_1X_2X'_3\rangle\,+\,\langle X_1X'_2X'_3\rangle\,-\,\langle X_1X_2X_3\rangle\,-\,\langle X'_1X'_2X'_3\rangle.
\end{split}
\end{equation}
Using Eq. (\ref{Z1}), $E_{\lambda}$ becomes,
\begin{equation}\label{Z3}
    E_{\lambda} \leq 8 - 2J
\end{equation}
Where,
%\begin{widetext}
\begin{equation}\label{Z4}
\begin{split}
&J = \sum_{i = 1}^{6} [|m_{i1} - m_{i2}| + |1 + m_{i1} + m_{i2} - m_{i4} - m_{i5} - m_{i6}|+ |m_{i4} + m_{i5} + m_{i6} - 2m_{i3} - 1 - |m_{i1} - m_{i2}|\\ &+ |1 + m_{i1} + m_{i2} - m_{i4} - m_{i5} - m_{i6}||]
+ \sum_{i = 7}^{8} [|m_{i2} + m_{i3} - m_{i4}| + |m_{i2} + m_{i6} - m_{i3} - m_{i5}| + |m_{i5} + m_{i6} -m_{i4} - 2m_{i1} \\
& + |m_{i2} + m_{i3} - m_{i4}| - |m_{i2} + m_{i6} - m_{i3} - m_{i5}||].
\end{split}
\end{equation}
%\end{widetext}
Due to indeterminism, any marginal probability must lie in $[0,\,I]$ or $[1-I,\,1]$. Due to signaling, $S_{ij\rightarrow k}> \,0; \, S_{i\rightarrow jk}> \,0,\:\forall i,j,k\,\in\{1,2,3\} \textmd{and}\: i\neq j\neq k$. Under these constraints, maximizing the quantity $E_{\lambda}$ corresponds to minimizing the quantity $J$.
To proceed, suppose first that $S < 1- 2I$. Using this criteria,
\begin{equation}\label{Z5}
\begin{split}
       J&\geq 2(1 - 3I) \qquad\qquad \text{for} \quad 0 \leq I \leq \frac{2}{9},\\
        &\geq 6(1 - 4I)      \qquad\qquad  \text{for} \quad \frac{2}{9} \leq  I < \frac{1}{4},\\
        &\geq 0             \qquad\qquad\qquad \quad \,\,\text{for}\quad \frac{1}{4} \leq  I < \frac{1}{2}.
\end{split}
\end{equation}
Equality is obtained via the choices: \{$ m_{i4} = I, m_{i6} = 1 - I, m_{ij}=0$ (for j\,$\neq $\,$4$,\,6; $i = 1, \ldots, 6$)
 and $m_{i2} = 0,\,m_{i6} = 1- I,\,m_{ij}=I\,$ ($j\,\neq 2, 6$;\, $i = 7, 8)$\} for $0 \leq I \leq \frac{2}{9}$, \{$m_{ij} = 0$ ($j=1, \ldots, 6$; $i =7, 8$), $m_{ij} = 0$ ($j\,\neq 4,5;\,i= 1, 4$),  $m_{14} = I, m_{15} = 1 - I, m_{45} = 0, m_{44} = 1$, $ m_{i4} = I, m_{i6} = 1- I, m_{ij}=0$ ($j\,\neq 4,6;\,i = 2, 6$), $m_{i6} = 1 - I,m_{ij} = I$  ($j=1,2,3;\, i = 3, 5$), $m_{34} = 0, m_{35} = 1 - I, m_{54} = 1 - I, m_{55} = 0$\} for $\frac{2}{9} \leq  I < \frac{1}{4}$ and \{$m_{ij}= 0$ ($j=1, \ldots, 6$; $i =7, 8$), $m_{ij} = \frac{1}{2} - I$(for $j=1,2,3; i= 1, \ldots, 6$), $m_{i6} = 0, m_{i4} = m_{i5} = 1 - I$ ($i= 1 , 4$), $m_{i5} = 0, m_{i4} = m_{i6} = 1 - I$ ($i = 2, 5$), $m_{i4} = 0$, $m_{i5} = m_{i6} = 1 - I$ ($i = 3, 6$) \} for $\frac{1}{4} \leq  I < \frac{1}{2}$. Thus, for $S < 1 - 2I $, Eq.(\ref{s8}) immediately follows via Eq.(\ref{Z3} and \ref{Z5}).
Finally, if $S \geq 1 - 2I$, then $J \geq 0 $ where equality holds for $m_{ij} = 0$ (for $i =\,7, 8,\,j=\,1, \ldots, 6$), $ m_{i4} = I,\, m_{i5} = 1 - I,\, m_{ij}=0 \,$(for j\,$\neq $\,4,5; $i = 1, \ldots, 6$). Hence Eq.(\ref{Z3}) yields the tight bound $E_{\lambda} \leq 8$ as desired. $\blacksquare$\\
\end{widetext}
\end{document}